\documentclass{webofc}
\usepackage[varg]{txfonts}
\usepackage{hyperref}
\usepackage{url}
\hypersetup{colorlinks=true,citecolor=blue,urlcolor=blue,linkcolor=blue}

\newcommand{\amu}[1]{a^{\text{HVP}(#1)}_{\mu}}
\newcommand{\APF}[1]{A_{0}^{(#1)}}
\newcommand{\KG}[2]{K_{#1}^{(#2)}}
\newcommand{\KGt}[2]{\tilde{K}_{#1}^{(#2)}}
\newcommand{\ind}[2]{^{#1}_{\text{#2}}}
\newcommand{\mmu}{m_{\mu}}
\newcommand{\iep}{i\varepsilon}
\newcommand{\DP}{\Delta\Pi}
\newcommand{\Nc}{N_{\text{c}}}
\newcommand{\nf}{n_{f}}

\newcommand{\arctanh}{{\rm arctanh}}

\allowdisplaybreaks
\begin{document}

\title{HVP~contributions to the muon anomalous magnetic moment
in spacelike and timelike domains}

\author{\firstname{Alexander} \lastname{Nesterenko}\inst{1}\fnsep\thanks{\email{nesterav@theor.jinr.ru}}}

\institute{Joint Institute for Nuclear Research, Dubna, 141980, Russian Federation}

\abstract{The hadronic vacuum polarization contributions to the muon
anomalous magnetic moment~$a^{\text{HVP}}_{\mu}$ in spacelike and timelike
domains are recapped. The values of~$a^{\text{HVP}}_{\mu}$ assessed within
dispersively-improved perturbation theory are reported. The obtained results
can be employed in the evaluations of~$a^{\text{HVP}}_{\mu}$ in the framework
of such methods as lattice studies, MUonE project, and others.}
\maketitle

\section{Introduction}
\label{Sect:Intro}

The theoretical description of the muon anomalous magnetic moment
\mbox{$a_{\mu} = (g_{\mu}-2)/2$} represents a long-standing challenging issue
of elementary particle physics, which engages all interactions of the
Standard Model. While the~experimental measurement of this quantity has
attained an impressive accuracy~\cite{Muong-2:2006rrc, Muong-2:2025xyk},
currently it is unclear whether it agrees or disagrees with its theoretical
evaluation due to substantial tensions between the values of the hadronic
vacuum polarization contribution~$a^{\text{HVP}}_{\mu}$ calculated within
lattice and timelike-data-driven methods and among those of the
timelike-data-driven method itself (for details, see a comprehensive
review~\cite{Aoyama:2020ynm} and its recent update~\cite{Aliberti:2025beg},
which is primarily based on Refs.~\cite{RBC:2018dos, Giusti:2019xct,
Borsanyi:2020mff, Lehner:2020crt, Wang:2022lkq, Aubin:2022hgm, Ce:2022kxy,
ExtendedTwistedMass:2022jpw, RBC:2023pvn, Kuberski:2024bcj,
Boccaletti:2024guq, Spiegel:2024dec, RBC:2024fic, Djukanovic:2024cmq,
ExtendedTwistedMass:2024nyi, MILC:2024ryz, FermilabLatticeHPQCD:2024ppc,
Keshavarzi:2019abf, DiLuzio:2024sps, Kurz:2014wya, Colangelo:2015ama,
Masjuan:2017tvw, Colangelo:2017fiz, Hoferichter:2018kwz, Eichmann:2019tjk,
Bijnens:2019ghy, Leutgeb:2019gbz, Cappiello:2019hwh, Masjuan:2020jsf,
Bijnens:2020xnl, Bijnens:2021jqo, Danilkin:2021icn, Stamen:2022uqh,
Leutgeb:2022lqw, Hoferichter:2023tgp, Hoferichter:2024fsj, Estrada:2024cfy,
Ludtke:2024ase, Deineka:2024mzt, Eichmann:2024glq, Bijnens:2024jgh,
Hoferichter:2024bae, Holz:2024diw, Cappiello:2025fyf, Colangelo:2014qya,
Blum:2019ugy, Chao:2021tvp, Chao:2022xzg, Blum:2023vlm, Fodor:2024jyn,
Aoyama:2012wk, Volkov:2019phy, Volkov:2024yzc, Aoyama:2024aly,
Parker:2018vye, Morel:2020dww, Fan:2022eto, Czarnecki:2002nt,
Gnendiger:2013pva, Hoferichter:2025yih}). In view of this, alternative
approaches, particularly a~novel spacelike-data-driven MUonE
method~\cite{CarloniCalame:2015obs, MUonE:2016hru}, gain an exceptional
significance, since they may shed additional light on this subject.

\section{Dispersion relations}
\label{Sect:DR}

The hadronic vacuum polarization function~$\Pi(q^2)$, being the
scalar part of the hadronic vacuum polarization tensor
\begin{equation}
\label{P_Def}
\Pi_{\mu\nu}(q^2) = i\!\int\!d^4x\,e^{i q x} \Bigl\langle 0 \bigl|\,
T\Bigl\{J_{\mu}(x)\, J_{\nu}(0)\Bigr\} \bigr| 0 \Bigr\rangle =
\frac{i}{12\pi^2} (q_{\mu}q_{\nu} - g_{\mu\nu}q^2) \Pi(q^2),
\end{equation}
is interrelated with the Adler function~$D(Q^2)$ and the so-called $R$-ratio
of electron-positron annihilation into hadrons in the following~way:
\begin{subequations}
\label{DispRels}
\begin{align}
\label{DispRelsP}
\DP(q^{2},\, q_{0}^{2}) & = (q^2 - q_0^2) \int_{s_{\text{th}}}^{\infty}
\frac{R(\sigma)}{(\sigma-q^2)(\sigma-q_0^2)}\, d\sigma =
- \int_{-q_0^2}^{-q^2} \!D(\sigma)\, \frac{d\sigma}{\sigma},
\\
\label{DispRelsD}
D(Q^2) & = - \frac{d\,\Pi(-Q^2)}{d\,\ln Q^2} =
Q^2\! \int_{s_{\text{th}}}^{\infty} \frac{R(\sigma)}{(\sigma+Q^2)^2}\, d\sigma,
\\
\label{DispRelsR}
R(s) & = \frac{1}{2 \pi i} \lim_{\varepsilon \to 0_{+}}
\Bigl[\Pi(s+i\varepsilon)-\Pi(s-i\varepsilon)\Bigr]\! =
\frac{1}{2 \pi i} \lim_{\varepsilon \to 0_{+}}
\int_{s+i\varepsilon}^{s-i\varepsilon} D(-\zeta)
\frac{d\zeta}{\zeta},
\end{align}
\end{subequations}
where \mbox{$Q^2 = -q^2 \ge 0$} and \mbox{$s = q^2 \ge 0$} denote,
respectively, the spacelike and timelike kinematic variables,
\mbox{$\DP(q^{2},\, q_{0}^{2}) = \Pi(q^2) - \Pi(q_0^2)$}, and $s_{\text{th}}$
stands for the hadronic production threshold, see
Refs.~\cite{Feynman:2019rot, Adler:1974gd, Radyushkin:1982kg,
Krasnikov:1982fx, Moorhouse:1976qq, Pennington:1981cw, Pennington:1983rz,
Pivovarov:1991bi}. A~detailed overview of this matter is given in, e.g.,
Chap.~1 of Ref.~\cite{Nesterenko:2016pmx} and references therein.

\section{Hadronic vacuum polarization contributions to~$a_{\mu}$}
\label{Sect:AmuHVP}

The HVP~contributions to the muon anomalous magnetic moment can be expressed
in terms of any of the functions~(\ref{DispRels}). Namely, in the
\mbox{$\ell$-th}~order in the electromagnetic coupling
\begin{subequations}
\label{Amu}
\begin{align}
\label{AmuP}
\amu{\ell} & = \APF{\ell}\int_{0}^{\infty}
\KG{\Pi}{\ell}(Q^2) \bar\Pi(Q^2) \frac{d Q^2}{4\mmu^2} =
\APF{\ell}\int_{0}^{\infty}
\KGt{\Pi}{\ell}(\zeta) \bar\Pi(4\zeta\mmu^2) d \zeta =
\\
\label{AmuD}
& = \APF{\ell}\int_{0}^{\infty}
\KG{D}{\ell}(Q^2) D(Q^2) \frac{d Q^2}{4\mmu^2} =
\APF{\ell}\int_{0}^{\infty}
\KGt{D}{\ell}(\zeta) D(4\zeta\mmu^2) d \zeta =
\\
\label{AmuR}
& = \APF{\ell}\int_{s_{\text{th}}}^{\infty}
\KG{R}{\ell}(s) R(s) \frac{d s}{4\mmu^2} =
\APF{\ell}\int_{\chi}^{\infty}
\KGt{R}{\ell}(\eta) R(4\eta\mmu^2) d \eta,
\end{align}
\end{subequations}
where~$\APF{\ell}$ is a constant prefactor, \mbox{$\bar\Pi(Q^2) = \DP(0,-Q^2)
= -\Pi(-Q^2)$} stands for the subtracted at zero hadronic vacuum polarization
function [with \mbox{$\Pi(0)=0$} being assumed], \mbox{$\zeta =
Q^2/(4\mmu^2)$} and~\mbox{$\eta = s/(4\mmu^2)$} are the dimensionless
kinematic variables, \mbox{$\chi=s_{\text{th}}/(4\mmu^2)$},
and~$\KG{\Pi}{\ell}(Q^2)$, $\KG{D}{\ell}(Q^2)$, $\KG{R}{\ell}(s)$ stand for
the corresponding kernels, which have been extensively studied over the past
decades in the~timelike domain, while their spacelike counterparts remained
mostly unavailable.

\subsection{Leading-order~$a^{\text{\normalfont HVP}}_{\mu}$}
\label{Sect:AmuHVP-LO}

\begin{figure}[t]
\centerline{%
\begin{tabular}{p{62.5mm}p{62.5mm}}
\centerline{\raisebox{2.25mm}{%
\includegraphics[height=38mm,clip]{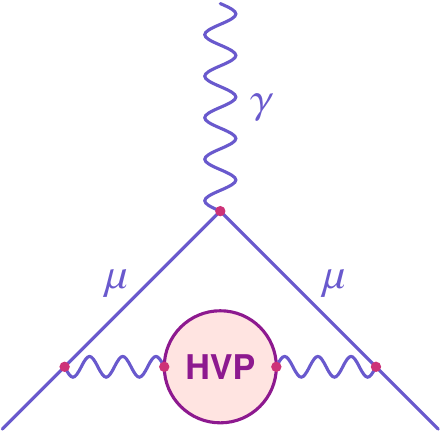}}}
&
\centerline{\includegraphics[width=62.5mm,clip]{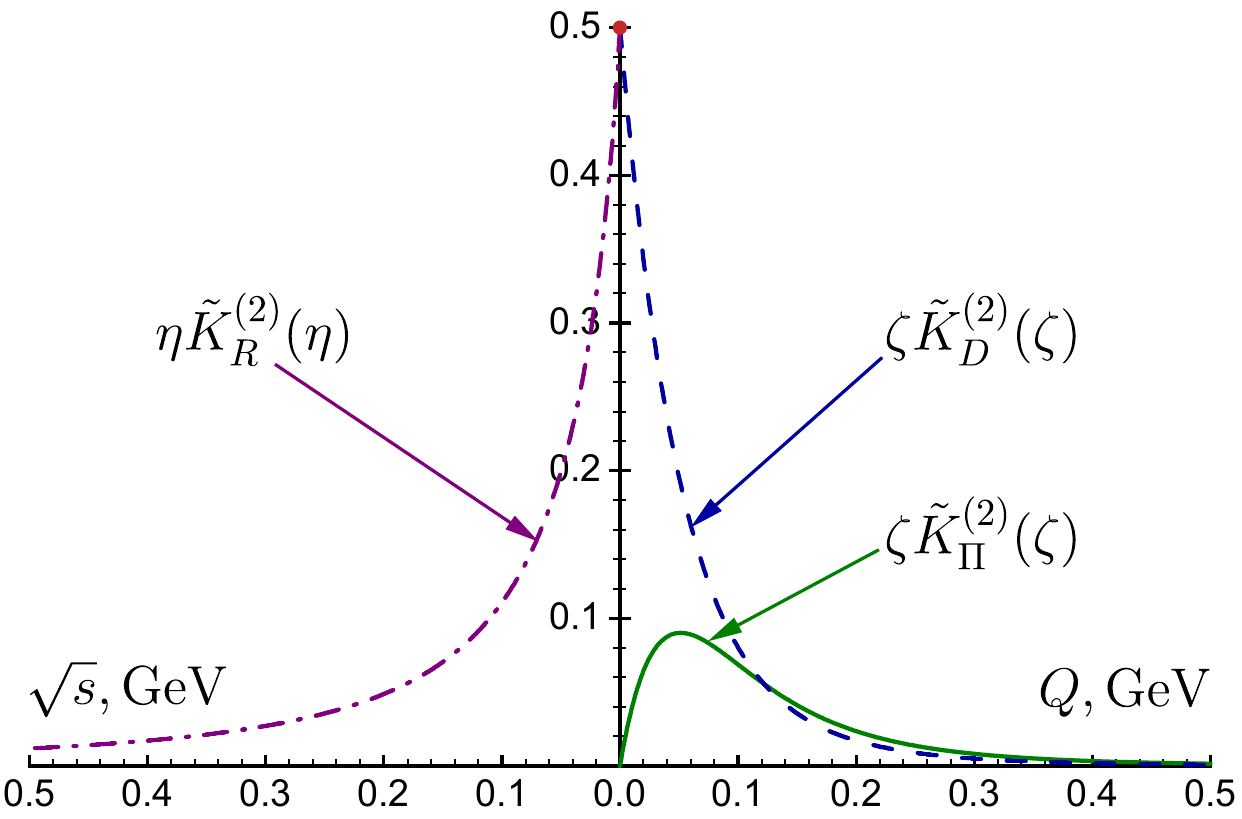}}
\end{tabular}}
\caption{Left~panel: leading-order hadronic vacuum polarization contribution
to the muon anomalous magnetic moment~(\ref{Amu2Def}). Right~panel:
leading-order kernels $\zeta\KGt{\Pi}{2}(\zeta)$ [Eq.~(\ref{KP2expl}), solid
curve], $\zeta\KGt{D}{2}(\zeta)$ [Eq.~(\ref{KD2expl}), dashed curve], and
$\eta\KGt{R}{2}(\eta)$ [Eqs.~(\ref{K2RInt}),~(\ref{K2RExpl}), dot--dashed
curve] in spacelike [\mbox{$Q^2 = -q^2 \ge 0$}, \mbox{$\zeta =
Q^2/(4\mmu^2)$}] and timelike [\mbox{$s = q^2 \ge 0$}, $\eta = s/(4\mmu^2)$]
domains.}
\label{Plot:AmuHVP2}
\end{figure}

\looseness=-1
In~the leading order of perturbation theory (that is, in the second order
in~the electromagnetic coupling,~$\ell=2$, see left panel of
Fig.~\ref{Plot:AmuHVP2}) a~peculiar form of the representation for the
timelike kernel~(\ref{K2RInt}) (see Refs.~\cite{Berestetskii:1956tkr,
Bouchiat:1961lbg, Kinoshita:1967txv}) allows one to express the
contribution~$\amu{2}$ in terms of the spacelike functions~$\bar\Pi(Q^2)$
(see Ref.~\cite{Lautrup:1971jf}) and~$D(Q^2)$~(see~Refs.~\cite{Knecht:2003kc,
deRafael:2017gay}) by making use of the dispersion
relations~(\ref{DispRels}), but (as~discussed in,
e.g.,~Refs.~\cite{deRafael:2017gay, Nesterenko:2021byp, Nesterenko:2023bdc})
only in such specific case:
\begin{align}
\label{Amu2Def}
\amu{2} & = \APF{2} \!\int_{s_{\text{th}}}^{\infty}\!
G_{R}^{(2)}\biggl(\frac{s}{4\mmu^2}^{\!}\biggr) R(s) \frac{ds}{s} =
\APF{2} \!\int_{0}^{1}\! (1-x)
\bar\Pi\biggl(\mmu^2\frac{x^2}{1-x}\biggr) d x
\nonumber \\ & =
\APF{2} \!\int_{0}^{1}\! (1-x) \biggl(\!1-\frac{x}{2}\biggr)
D\biggl(\mmu^2\frac{x^2}{1-x}\biggr) \frac{d x}{x},
\qquad
\APF{2} = \frac{1}{3} \biggl(\frac{\alpha}{\pi}\biggr)^{2},
\end{align}
where
\begin{equation}
\label{K2RInt}
K_{R}^{(2)}(4\eta\mmu^2) = \tilde{K}_{R}^{(2)}(\eta),
\qquad
G_{R}^{(2)}(\eta) = \eta\KGt{R}{2}(\eta) =
\int_{0}^{1}\! \frac{x^2 (1-x)}{x^2 + 4\eta (1-x)}\,dx.
\end{equation}

The leading-order spacelike kernels~$K_{\Pi}^{(2)}(Q^2)$
(see Refs.~\cite{Groote:2001vu, Blum:2002ii, Nesterenko:2014txa,
deRafael:2017gay}) and~$K_{D}^{(2)}(Q^2)$ (see Refs.~\cite{Groote:2001vu,
deRafael:2017gay}) can then be obtained by mapping the integration
range~\mbox{$0 \le x < 1$} in~Eq.~(\ref{Amu2Def}) onto the kinematic
interval~\mbox{$0 \le Q^2 < \infty$}, while the integration in
Eq.~(\ref{K2RInt}) for the timelike kernel can be performed explicitly (see
Refs.~\cite{Berestetskii:1956tkr, Durand:1962zzb, Brodsky:1967sr,
Lautrup:1968tdb}), namely
\begin{subequations}
\label{K2expl}
\begin{align}
\label{KP2expl}
&\!\! K_{\Pi}^{(2)}(4\zeta\mmu^2) = \tilde{K}_{\Pi}^{(2)}(\zeta),
\quad\,\,
G_{\Pi}^{(2)}(\zeta) =
\zeta\KGt{\Pi}{2}(\zeta) = \frac{1}{\zeta^2} \,
\frac{y^{5}(\zeta)}{1-y(\zeta)},
\quad\,\,
y(\zeta) = \zeta \Bigl(\!\sqrt{1+\zeta^{-1}}-1\Bigr),
\\
\label{KD2expl}
&\!\! K_{D}^{(2)}(4\zeta\mmu^2) = \tilde{K}_{D}^{(2)}(\zeta),
\qquad
G_{D}^{(2)}(\zeta) =
\zeta\KGt{D}{2}(\zeta) = (2\zeta+1)^{2}
- 2(2\zeta+1)\sqrt{\zeta(\zeta+1)} - \frac{1}{2},
\\
\label{K2RExpl}
&\!\! G_{R}^{(2)}(\eta) =
\!\frac{1}{2} + 4\eta\Bigl[(2\eta-1)\ln(4\eta)-1\Bigr]
-2\Bigl[2(2\eta-1)^2-1\Bigr]
\frac{\arctanh\bigl[\psi(\eta)\bigr]}{\psi(\eta)},
\quad\,
\psi(\eta) = \!\frac{\sqrt{\eta-1}}{\sqrt{\eta}}.
\end{align}
\end{subequations}
The right panel of Fig.~\ref{Plot:AmuHVP2} displays the
kernels~(\ref{K2expl}) in spacelike and timelike domains.

\subsection{Higher-order~$a^{\text{\normalfont HVP}}_{\mu}$}
\label{Sect:AmuHVP-HO}

Factually, since the functions~$\bar\Pi(Q^2)$, $D(Q^2)$, $R(s)$ entering the
representations for~$a^{\text{HVP}}_{\mu}$~(\ref{Amu}) are all
interrelated~(\ref{DispRels}), the corresponding kernels can also be
expressed in terms of each other. Specifically, a complete set of six
relations between the spacelike and timelike kernels $\KG{\Pi}{\ell}(Q^2)$,
$\KG{D}{\ell}(Q^2)$, $\KG{R}{\ell}(s)$ was obtained in
Ref.~\cite{Nesterenko:2021byp}, thereby enabling one to directly calculate
the unknown kernels by making use of a known one:
\begin{subequations}
\label{KRels}
\begin{align}
\label{KRelsP}
% \label{KRelPR}
\KG{\Pi}{\ell}(Q^2) & = \frac{1}{2 \pi i} \lim_{\varepsilon \to 0_{+}}
\Bigl[ \KG{R}{\ell}(-Q^2+\iep) - \KG{R}{\ell}(-Q^2-\iep)\Bigr] =
% \\
% \label{KRelPD} & =
- \Biggl[\KG{D}{\ell}(Q^2)
+ \frac{d\,\KG{D}{\ell}(Q^2)}{d\,\ln Q^2}\Biggr],
\\
\label{KRelsD}
% \label{KRelDP}
\KG{D}{\ell}(Q^2) & = \frac{1}{Q^2}
\int_{Q^2}^{\infty} \KG{\Pi}{\ell}(p^{2})\, d p^{2}
=
% \frac{4\mmu^{2}}{Q^2}K_{0}^{(\ell)} -
% \frac{1}{Q^2}
% \int_{0}^{Q^2} \KG{\Pi}{\ell}(\xi)\, d \xi =
% \\
% \label{KRelDR} & =
- \frac{1}{2 \pi i} \lim_{\varepsilon \to 0_{+}}
\frac{1}{Q^2}
\int_{Q^2 + \iep}^{Q^2 - \iep}
\KG{R}{\ell}(-p^2) d p^2,
\\
\label{KRelsR}
% \label{KRelRP}
\KG{R}{\ell}(s) & = \frac{1}{s} \int_{0}^{\infty}
\KG{\Pi}{\ell}(Q^2) \frac{Q^2}{s + Q^2}\, d Q^2 =
% \\
% \label{KRelRD} & =
\int_{0}^{\infty}
\KG{D}{\ell}(Q^2) \frac{Q^2}{(s + Q^2)^2}\, d Q^2
\end{align}
\end{subequations}
(note that the first of the relations~(\ref{KRelsP}) has also been
independently derived in a different way in~Ref.~\cite{Balzani:2021del}).
Additionally, the first equalities of Eqs.~(\ref{KRelsD}) and~(\ref{KRelsR})
imply that the functions~$Q^{2} K_{D}(Q^2)$ and~$s K_{R}(s)$ have the same
infrared limiting value~\cite{Nesterenko:2021byp}:
\begin{equation}
\label{KRDlim}
K_{0}^{(\ell)} =
\lim_{Q^2 \to 0_{+}} \frac{Q^2}{4\mmu^{2}}\,\KG{D}{\ell}(Q^2)
=
\lim_{s \to 0_{+}} \frac{s}{4\mmu^{2}}\, \KG{R}{\ell}(s)
=
\int_{0}^{\infty}\!\KG{\Pi}{\ell}(p^{2})\, \frac{d p^{2}}{4\mmu^{2}}.
\end{equation}
In~particular, it is straightforward to verify that the leading-order
kernels~(\ref{K2expl}) explicitly satisfy all the relations~(\ref{KRels}),
while the infrared limiting value~(\ref{KRDlim}), namely~$K_{0}^{(2)}=1/2$,
corresponds to the leading Schwinger contribution~\cite{Schwinger:1948iu},
see Refs.~\cite{Nesterenko:2021byp, Nesterenko:2023bdc} for the details.

As~outlined earlier, the way the spacelike kernels~(\ref{KP2expl})
and~(\ref{KD2expl}) were derived from the timelike expression~(\ref{Amu2Def})
relies solely on a specific form of the leading-order kernel~(\ref{K2RInt})
and cannot be performed in a general case. Nonetheless, the obtained
relations~(\ref{KRels}) enable one to directly calculate the unknown kernels
by making use of a~known one, that significantly facilitates the derivation
of the spacelike kernels beyond the leading order, see
Refs.~\cite{Nesterenko:2021byp, Balzani:2021del}.

\section{Updated DPT value of~$a_{\mu}$}
\label{Sect:AmuDPT}

Since the involved energy scales are essentially low, one inevitably needs to
resort to either a data-driven method or a nonperturbative approach herein.
In what follows the dispersively-improved perturbation theory (DPT)
\cite{Nesterenko:2013vja, Nesterenko:2014txa, Nesterenko:2016pmx}, which
self-consistently combines the nonperturbative constraints imposed by the
dispersion relations (\ref{DispRels}) on the functions $\bar\Pi(Q^2)$,
$D(Q^2)$, $R(s)$ with corresponding perturbative input, will be employed (see
also Refs.~\cite{Nesterenko:2005wh, Nesterenko:2007fm, Nesterenko:2008fb}).
Specifically, the DPT enables one to overcome some inherent difficulties of
the QCD perturbation theory and extends its applicability range towards the
infrared domain, see Refs.~\cite{Nesterenko:2013vja, Nesterenko:2014txa,
Nesterenko:2016pmx} for the details. In particular, the DPT hadronic vacuum
polarization function, being applicable in the whole energy range, makes it
possible to directly employ it in the corresponding expression for
$a^{\text{HVP}}_{\mu}$~(\ref{AmuP}) without using timelike data on
\mbox{$R$-ratio}~\cite{Nesterenko:2014txa, Nesterenko:2015rea,
Nesterenko:2017hqv, Nesterenko:2023bdc, Nesterenko:2025mou}.
\looseness=-1

The~DPT provides the following integral representation for the hadronic
vacuum polarization function~\cite{Nesterenko:2013vja, Nesterenko:2014txa,
Nesterenko:2016pmx}:
\begin{equation}
\label{DPT:P}
\DP(q^2,\, q_0^2) = \Nc\!\sum_{f=1}^{\nf}\! Q_{f}^{2} \left[
\DP^{(0)}(q^2,\, q_0^2) +
\!\int_{s_{\text{th}}}^{\infty}\! \rho(\sigma)
\ln\Biggl(\frac{\sigma-q^2}{\sigma-q_0^2}
\frac{s_{\text{th}}-q_0^2}{s_{\text{th}}-q^2}\Biggr)\frac{d\sigma}{\sigma}
\right]\!,
\end{equation}
where \mbox{$\Nc=3$} denotes the number of colors, $Q_{f}$~is the electric
charge of the quark of $f$-th flavor (in units of the elementary charge),
$\nf$~stands for the number of active flavors, the~leading-order term
reads~\cite{Feynman:2019rot, Akhiezer:1965}
\begin{equation}
\label{DPT:P0L}
\DP^{(0)}(q^2, q_0^2) = 2_{\,}\Biggl(\!1_{\!}+_{\!}\frac{1}{\xi}\Biggr)
\biggl[1-\!\!\sqrt{1+\xi^{-1}}\,\text{arcsinh}\!\left(\xi^{1/2}\right)\biggr]
-2_{\,}\Biggl(\!1_{\!}+_{\!}\frac{1}{\xi_{0}}\Biggr)
\biggl[1-\!\!\sqrt{1+\xi_{0}^{-1}}\,\text{arcsinh}\!\left(\xi_{0}^{1/2}\right)\biggr],
\;
\end{equation}
$\xi=-q^2/s_{\text{th}}$, $\xi_{0}=-q_{0}^{2}/s_{\text{th}}$, while the
spectral function provides the corresponding perturbative input, specifically
\begin{equation}
\label{RhoPert}
\rho(\sigma) =
\rho\ind{}{pert}(\sigma) = \frac{1}{\pi} \frac{d}{d\,\ln\sigma}\,
\text{Im}\, p\ind{}{pert}(\sigma-i0_{+}) =
- \frac{d\, r\ind{}{pert}(\sigma)}{d\,\ln\sigma}
= \frac{1}{\pi}\, \text{Im}\,d\ind{}{pert}(-\sigma-i0_{+}),
\end{equation}
with $p\ind{}{pert}(q^2)$, $r\ind{}{pert}(s)$, and~$d\ind{}{pert}(Q^2)$ being
the perturbative strong corrections to the functions~$\Pi(q^2)$, $R(s)$,
and~$D(Q^2)$, respectively, see Refs.~\cite{Nesterenko:2013vja,
Nesterenko:2014txa, Nesterenko:2016pmx} and references therein for the
details. In~the first order in the strong coupling the perturbative spectral
function~(\ref{RhoPert}) takes a~simple form, namely,
$\rho\ind{(1)}{pert}(\sigma) = (4/\beta_{0})
[\ln^{2}(\sigma/\Lambda^2)+\pi^2]^{-1}$, where $\beta_{0} = 11 - 2\nf/3$
stands~for the one-loop $\beta$~function perturbative expansion coefficient
and $\Lambda$~is the QCD scale parameter, whereas the explicit expression for
the higher-loop perturbative spectral function~(\ref{RhoPert}) was obtained
in Refs.~\cite{Nesterenko:2016pmx, Nesterenko:2017wpb}, see
also~Refs.~\cite{Nesterenko:2018pks, Nesterenko:2019rag}.

In~the leading and in the next-to-leading orders of perturbation theory
[that~is, in~the second~(\mbox{$\ell=2$}) and in~the third~(\mbox{$\ell=3$})
orders in~the electromagnetic coupling] the DPT hadronic vacuum polarization
function~(\ref{DPT:P}) taken in the fourth order in the strong coupling
yields~\cite{Nesterenko:2025mou}
\begin{equation}
\label{Amu2DPT25}
\amu{2} = (695.95 \pm 2.83)\times 10^{-10},
\end{equation}
\vskip-7.5mm
\begin{alignat}{2}
\label{Amu3DPT25}
\amu{3a} &= (-218.08 \pm 0.96)\times 10^{-11},
&\qquad
a^{\text{HVP}(3b)}_{\mu, e} &= (106.53 \pm 0.44)\times 10^{-11},
\nonumber \\*[1mm]
a^{\text{HVP}(3b)}_{\mu, \tau} &= (559.67 \pm 2.08)\times 10^{-15},
&\qquad
\amu{3c} &= (334.92 \pm 2.66)\times 10^{-13},
\end{alignat}
that agrees with the timelike-data-driven evaluations~\cite{Aoyama:2020ynm,
Davier:2019can, Keshavarzi:2019abf, Kurz:2014wya, Jegerlehner:2017lbd}.
Equations~(\ref{Amu2DPT25}) and~(\ref{Amu3DPT25}) correspond to
the~values~\cite{ParticleDataGroup:2024cfk} of the involved Standard Model
parameters, including the world average~$\alpha_{\text{s}}(M_{Z}^{2})$, while
the employed definitions and notations can be found in
Refs.~\cite{Nesterenko:2021byp, Nesterenko:2023bdc}. Combining the
values~(\ref{Amu2DPT25}) and~(\ref{Amu3DPT25}) with the remaining
contributions~$a^{\text{QED}}_{\mu}$, $a^{\text{EW}}_{\mu}$, $\amu{4}$,
$a^{\text{HLbL}}_{\mu}$ reported~in Ref.~\cite{Aliberti:2025beg}
results~in~\cite{Nesterenko:2025mou}
\begin{equation}
\label{AmuSMDPT25}
a^{\text{SM}}_{\mu} = (11659185.25 \pm 3.00)\times 10^{-10},
\end{equation}
that constitutes an update of the assessments~\cite{Nesterenko:2014txa,
Nesterenko:2015rea, Nesterenko:2017hqv, Nesterenko:2023bdc}.

\begin{figure}[t]
\centerline{%
\raisebox{3.52mm}{%
\includegraphics[width=54mm,clip]{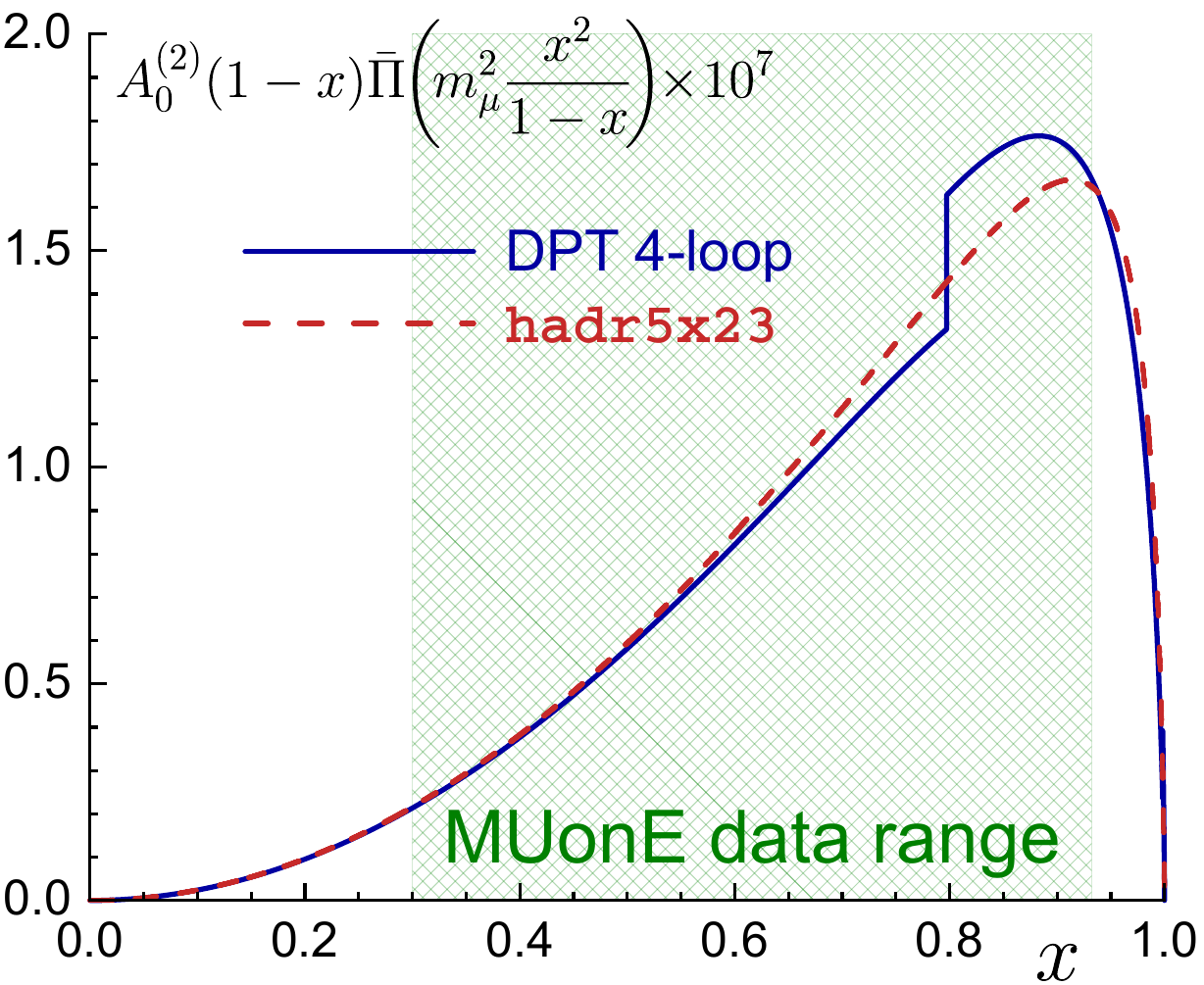}}%
\hspace{11mm}%
\includegraphics[width=54mm,clip]{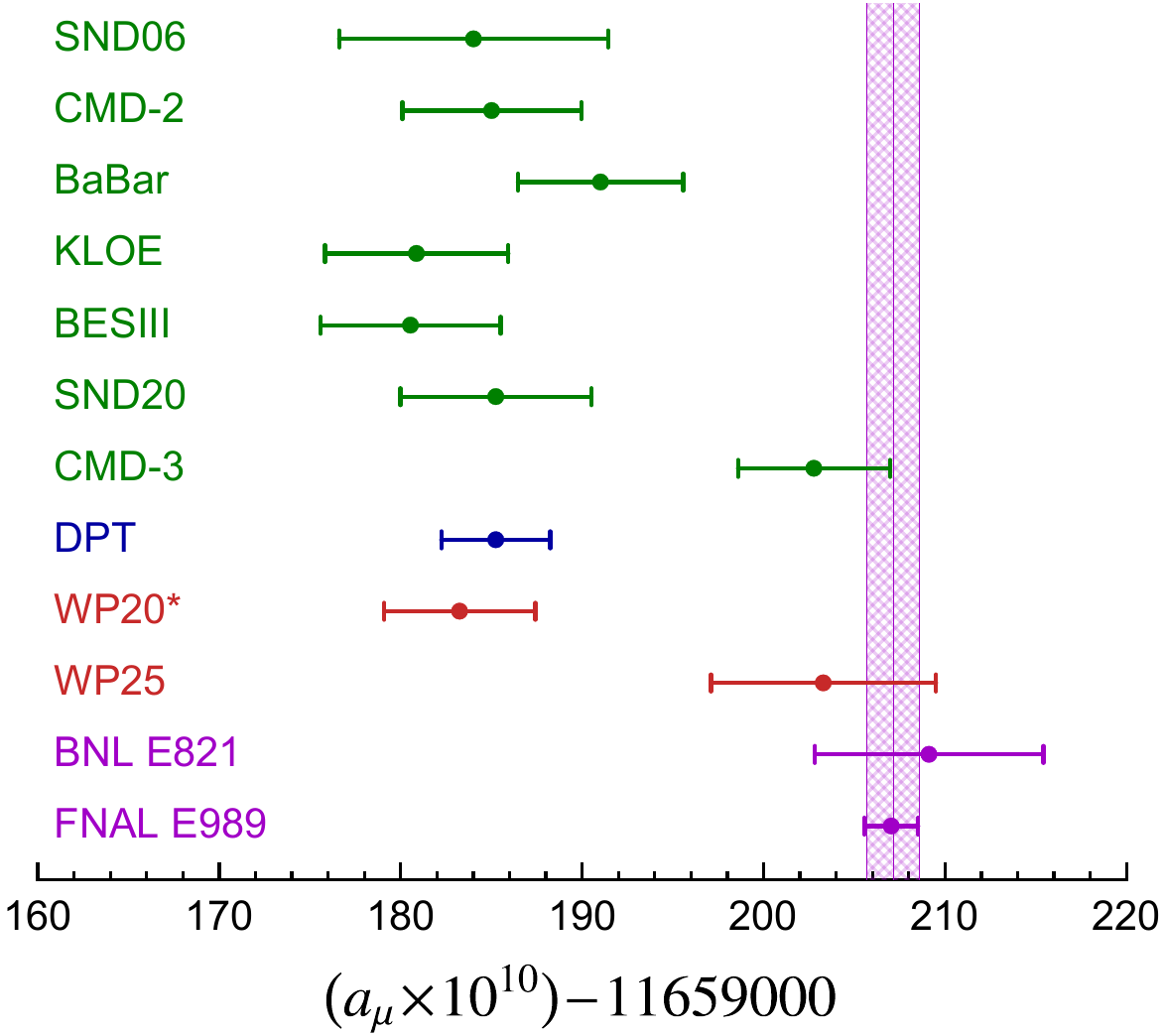}}
\caption{Left~panel: integrands of the second equality in Eq.~(\ref{Amu2Def})
corresponding to the DPT hadronic vacuum polarization function~(\ref{DPT:P})
(blue solid curve) and to the timelike-data-driven one computed with
the~\texttt{hadr5x23} routine~\cite{FJhadr5x23} (red dashed curve).
Right~panel: summary of the muon anomalous magnetic moment
assessments~\cite{Aliberti:2025beg} (labels SND06, CMD-2, BaBar, KLOE,
\mbox{BESIII}, SND20, CMD-3, WP20$^{*}$,~WP25), its
DPT~evaluation~(\ref{AmuSMDPT25}), experimental measurements
BNL~E821~\cite{Muong-2:2006rrc} and FNAL~E989~\cite{Muong-2:2025xyk}, and
their~combined result~(\ref{AmuSMexp25}) (vertical band).}
\label{Plot:Amu25}
\end{figure}

Figure~\ref{Plot:Amu25} (left~panel) shows the integrands of the second
equality in Eq.~(\ref{Amu2Def}) corresponding to the DPT~hadronic vacuum
polarization function~(\ref{DPT:P}) (blue solid curve) and to the
timelike-data-driven one computed with the~\texttt{hadr5x23}
routine~\cite{FJhadr5x23} (red dashed curve). The~latter parameterizes the
function~$R(s)$ appearing in the dispersion relation~(\ref{DispRelsP}) by the
experimental data on the $R$-ratio of electron-positron annihilation into
hadrons at low energies and by the respective perturbative expression at high
energies, that eventually smoothes out the quark flavor thresholds and
hadronic resonances. As one can infer from the left~panel
of~Fig.~\ref{Plot:Amu25}, the DPT hadronic vacuum polarization
function~(\ref{DPT:P}), being in an excellent agreement with its experimental
prediction at low energies, can also be employed as a supplementing infrared
input for the lattice studies (see~e.g.,~reviews~\cite{Meyer:2018til,
Wittig:2023pcl}) as well as for the MUonE
project~\cite{CarloniCalame:2015obs, MUonE:2016hru} in the energy range
uncovered by measurements, see Refs.~\cite{Nesterenko:2023bdc,
Nesterenko:2025mou} for the details.

Figure~\ref{Plot:Amu25} (right~panel) displays a summary of the muon
anomalous magnetic moment evaluations~\cite{Aliberti:2025beg}. The~datapoints
labeled SND06, CMD-2, BaBar, KLOE, \mbox{BESIII}, SND20, CMD-3 include all
the contributions to~$a_{\mu}$ reported in Ref.~\cite{Aliberti:2025beg},
except for~$\amu{2}$, which is assessed as~follows. The specified
measurements of \mbox{$e^{+}e^{-} \to \pi^{+}\pi^{-}$} mode below~$1.8$~GeV
were processed within CHKLS, DHMZ, and KNTW approaches (see~Sect.~2.6.1,
Sect.~2.5.1, and Sect.~2.5.2 of Ref.~\cite{Aliberti:2025beg}, respectively)
and then the averaged results were combined with the contributions from other
modes and from the energies above~$1.8$~GeV taken from
Ref.~\cite{Aoyama:2020ynm}, see Sects.~2 and~9 of
Ref.~\cite{Aliberti:2025beg} for the details. The~measurements SND06, CMD-2,
BaBar, KLOE, \mbox{BESIII} entered the result~\cite{Aoyama:2020ynm}
(similarly, the datapoint labeled WP20$^{*}$ combines~$\amu{2}$ from
Ref.~\cite{Aoyama:2020ynm} with the other contributions to~$a_{\mu}$ from
Ref.~\cite{Aliberti:2025beg}), whereas SND20 and \mbox{CMD-3} appeared
later~on. Because~of a yet-unresolved tension between CMD-3 and the other
experimental measurements, the timelike-data-driven contribution~$\amu{2}$
was not included in the assessment~of~$a_{\mu}$ reported in
Ref.~\cite{Aliberti:2025beg} (the~datapoint labeled~WP25), that uses the
lattice evaluation of~$\amu{2}$ instead, see Ref.~\cite{Aliberti:2025beg} for
the details. The vertical band represents the combined result of
BNL~E821~\cite{Muong-2:2006rrc} and FNAL~E989~\cite{Muong-2:2025xyk}
measurements
\begin{equation}
\label{AmuSMexp25}
a^{\text{exp}}_{\mu} = (11659207.15 \pm 1.45)\times 10^{-10}.
\end{equation}
As~one can infer from the right panel of Fig.~\ref{Plot:Amu25}, the
DPT~evaluation of~$a_{\mu}$~(\ref{AmuSMDPT25}) (the~datapoint labeled~DPT)
agrees with most of the timelike-data-driven assessments.

\section{Conclusions}
\label{Sect:Concl}

The obtained complete set of six relations~(\ref{KRels}) between the
spacelike and timelike kernels for~$a^{\text{HVP}}_{\mu}$~(\ref{Amu}) enables
one to directly calculate the unknown kernels by making use of a~known one,
that substantially facilitates the derivation of the spacelike kernels beyond
the leading order. Evaluation of the muon anomalous magnetic moment within
dispersively-improved perturbation theory~(\ref{AmuSMDPT25}) agrees with most
of the timelike-data-driven assessments. Additionally,
the~dispersively-improved perturbation theory provides a~supplementing
infrared input for the spacelike methods. The~obtained results can be
employed in the evaluations of~$a^{\text{HVP}}_{\mu}$ in the framework of
such methods as lattice studies, MUonE project, and others.

\bibliography{paper}

\end{document}